\documentstyle[12pt]{article}
\newcommand{\beq}{\begin{equation}}
\newcommand{\eeq}{\end{equation}}
\newcommand{\beqa}{\begin{eqnarray}}
\newcommand{\eeqa}{\end{eqnarray}}
\newcommand{\nn}{\nonumber}

\newcommand{\eq}[1]{(\ref{#1})}
\newcommand{\bd}{b^\dagger}
\newcommand{\ad}{a^\dagger}
\begin{document}

\begin{tabbing}
\` OS-GE-36-93 (revised) \\
\` November 1993
\end{tabbing}
\begin{center}
\vfill
{\large Landau Levels and Quantum Group}

\vspace{2cm}

Haru-Tada Sato$^{\dagger}$
\vspace{0.5cm}

{\em Institute of Physics, College of General Education}

{\em Osaka University, Toyonaka, Osaka 560, Japan}

\end{center}

\vfill
\begin{abstract}
We find a quantum group structure in two-dimensional motions
of a nonrelativistic electron in a uniform magnetic field
and in a periodic potential. The representation basis of the quantum
algebra is composed of wavefunctions of the system. The quantum group
symmetry commutes with the Hamiltonian and is relevant to the Landau
level degeneracy. The deformation parameter $q$ of the quantum algebra
turns out to be given by the fractional filling factor $\nu=1/m$
($m$ odd integer).
\end{abstract}
\vspace{1cm}
%
\noindent-----------------------------------------------------------------------
$^{\dagger}$ {\footnotesize Fellow of the Japan Society for the
Promotion of Science} \\
\mbox{}\hspace{0.3cm}{\footnotesize E-mail address~: hsato@jpnyitp.
yukawa.kyoto-u.ac.jp}
\newpage
%
%
 There have been many discussions in the study of quantum groups and
algebras \cite{QG,QG2}. Quantum group structures are found in
(2+1)-dimensional topological Chern-Simons theories \cite{CS} as well
as in rational conformal field theories and integrable lattice models
\cite{YB}. Although the abelian Chern-Simons theory does not possess a
quantum group structure in the literature \cite{CS}, it might be
possible to exhibit one in some other senses. There have been also
interesting investigations of condensed matter problems such as the
fractional quantum Hall effect making use of the abelian Chern-Simons
theory \cite{QH}. These observations bring us to find a quantum group
structure in two-dimensional motion of nonrelativistic electrons in a
uniform magnetic field.

 In this paper we consider the one body problem of such electron system
and derive a quantum group algebra acting within each Landau level. It
is also shown that the presence of a periodic potential term gives rise
to the same quantum group structure. These quantum group structures
allow us to expect a new approach to the quantized Hall effect utilizing
the representation theory of the quantum groups.

 Let us review some basic facts about non-interacting charged particle
(charge $e$ and mass $m$) in a constant magnetic field $B$ perpendicular
to the $x$-$y$ plane \cite{VO}. It is useful to adopt the following
phase space variables instead of $x_i$ and $p_i$:
\beqa
& \pi_i = p_i -{e\over c}A_i   \label{eq1}   \\
& \beta_i = \pi_i - m\omega\epsilon_{ij}x^j \label{eq2}
\eeqa
where $\epsilon_{11}=\epsilon_{22}=0$, $\epsilon_{12}=-\epsilon_{21}=1$
and
\beqa
& \omega = {eB\over mc} \label{3}   \\
& A_i = -{1\over2}B\epsilon_{ij}x^j -\partial_i\Lambda. \label{4}
\eeqa
The vector $\beta$ is related to the cyclotron center and the scalar
function $\Lambda$ is the gauge function which will be fixed later.
The commutation relations at the quantum level are
\beqa
&[\pi_i,\pi_j]=[\beta_j,\beta_i]=i\hbar m\omega\epsilon_{ij} \nn \\
&\hspace{1cm} [\pi_i,\beta_j]=0 \label{5}
\eeqa
and thus $\beta_i$ commute with the Hamiltonian
\beq
H={1\over 2m}(\pi_1^2 + \pi_2^2 ). \label{eq6}
\eeq
The diagonalized angular momentum is given by
\beq
J_3=\hbar(\bd b -\ad a) \label{eq7}
\eeq
where
\beqa
&\left(\begin{array}{c}
a \\
\ad
\end{array} \right)={1\over (2m\hbar\omega)^{1/2}}
(\pi_1 \pm i\pi_2) \label{eq8} \\
&\left(\begin{array}{c}
b \\
\bd
\end{array}\right)={1\over (2m\hbar\omega)^{1/2}}
(\beta_2 \pm i\beta_1) \label{eq9}
\eeqa
and these satisfy
\beq
[a,\ad]=[b,\bd]=1  \label{eq10}
\eeq
and all other commutators are zero. The Hamiltonian being reexpressed as
\beq
H=\hbar\omega(\ad a +{1\over2}), \label{eq11}
\eeq
it is obvious that $J_3$ commutes with the Hamiltonian. The dynamical
symmetry $su(2)$ is thus generated by
\beq
j_+=\bd a,\hskip 20pt j_-=\ad b,\hskip 20pt j={1\over2}(\bd b-\ad a).
\label{EX}
\eeq
It is worth noticing that the polynomials \cite{CTZ}
\beq
W_n^m=(\beta_1)^{n+1}(\beta_2)^{m+1},\hskip 20pt n,m\geq-1\label{eq13}
\eeq
satisfy the $W_{\infty}$ algebra \cite{Winf}
\beq
[W_n^m,W_k^l]=i\hbar m\omega\{(m+1)(k+1)-(n+1)(l+1)\}W_{n+k}^{m+l}
              +O(\hbar^2)\label{eq14}
\eeq
and
\beq [W_n^m,H]=0.\label{eq15} \eeq
Combining the generators $W_n^m$ in the following way
\beq
T_{(\alpha_1,\alpha_2)}=exp(-i{m\omega\over\hbar}
{\alpha_1\alpha_2\over2})\sum_{n,m=0}^{\infty}
({i\over\hbar})^{n+m}\alpha_1^n\alpha_2^mW_{n-1}^{m-1},\label{eq16}
\eeq
we get the magnetic translation operators \cite{mag}
\beq
T_{\bf \alpha}=exp({i\over\hbar}{\bf \alpha}\cdot{\bf \beta}),
\hskip 20pt {\bf \alpha}=(\alpha_1,\alpha_2) \label{eq17}
\eeq
which are interpreted as the translations accompanied by the gauge
transformations;
\beq
T_{\bf \alpha}=exp\left({i\over2}l_B^{-2}\epsilon_{ij}x^i
 \alpha^j+{ie\over\hbar c}\{\Lambda({\bf x}+{\bf\alpha})
-\Lambda({\bf x}) \} \right)\tau_{\bf\alpha} \label{eq18}
\eeq
where
\[
l_B=\sqrt{ {\hbar \over m\omega} }\,,\hskip 25pt
\tau_{\bf \alpha}=exp({i\over\hbar}{\bf \alpha}\cdot{\bf p}).
\]
The $l_B$ is called the magnetic length which describes the radius of
the area occupied by a degenerate state. The commutation relations
among the magnetic translations become (for example see \cite{Mac})
\beq
[T_{\bf a},T_{\bf b}]=2i\sin({1\over2}l_B^{-2}\epsilon_{ij}a^i b^j)
T_{{\bf a}+{\bf b}}. \label{EA}
\eeq

Now we show a quantum algebra ${\cal U}_q(su(2))$ of which deformation
parameter is defined by physical quantities. Let us introduce the
following combinations of the magnetic translations which translate a
fundamental distance $\Delta$ yet to be determined later:
\beq
{\cal L}_{-1}={T_{(-\Delta, -\Delta)}-T_{(\Delta, -\Delta)}
              \over q-q^{-1}} \label{ED}
\eeq

\beq
{\cal L}_{1}={T_{(\Delta, \Delta)}-T_{(-\Delta, \Delta)}
             \over q-q^{-1}} \label{EE}
\eeq

\beq
{\cal K}=T_{(2\Delta,0)}. \label{EF}
\eeq
These operators satisfy the following relations
\beq
[{\cal L}_1,{\cal L}_{-1}]
={{\cal K}-{\cal K}^{-1} \over q-q^{-1}}\,\,, \hskip 30pt
{\cal K}{\cal L}_{\pm1}{\cal K}^{-1}=q^{\pm2}{\cal L}_{\pm1}\label{EH}
\eeq
with the identification
\beq
 q=exp(i\Delta^2 l_B^{-2})\,. \label{EI}
\eeq
The algebra \eq{EH} is nothing but the quantum algebra
${\cal U}_q(su(2))$ identifying
$E_{+}={\cal L}_1$, $E_{-}={\cal L}_{-1}$ and $t={\cal K}$;
\beq
[E_{+},E_{-}]={t-t^{-1} \over q-q^{-1}}\,,\hskip 30pt
 tE_{\pm}t^{-1}=q^{\pm2}E_{\pm}\,. \label{eq88}
\eeq

We then consider a representation of the quantum algebra in order to
determine $\Delta$. The operator ${\cal K}$ and the Hamiltonian
commute with each other and are simultaneously diagonalizable. Choosing
the gauge $\Lambda={1\over2}Bxy$ for simplicity, the simultaneous
eigenfunction for both is found to be exactly the familiar Landau
states \cite{LL}
\beq
\vert n,l\rangle\equiv exp\{ 2\pi i{l\over L_x}x-{1\over 2l_B^2}(y-y_0)^2 \}
H_n({y-y_0 \over l_B}) \label{EJ}
\eeq
and
\beq
y_0=-2\pi l_B^2 {l\over L_x} \label{eq35}
\eeq
on which we have imposed the periodic boundary condition;
\beq
p_x=2\pi\hbar{l \over L_x}\,. \label{eq36} \eeq
$H_n(x)$ is the Hermite polynomial. We have ignored
ortho-normalization factor for \eq{EJ}, which is not important in
the following argument. From eqs.\eq{ED}, \eq{EE} and
\eq{EJ}, we see
\[
{\cal L}_{\pm}\vert n,l\rangle=[{1\over2}\pm2\pi l{l_B^2\over L_x\Delta}]
\vert n,l\pm{L_x\over 2\pi l_B^2}\Delta\rangle\,,\]
where the notation $[x]$ means $[x]={q^x-q^{-x}\over q-q^{-1}}$
and it becomes $x$ in the limit $q\rightarrow 1$.
If we require that the representation space is spanned by all the
degenerate Landau states \eq{EJ}, we should choose
$L_x\Delta/2\pi l_B^2=1$. Hence
\beq
\Delta=2\pi {l_B^2 \over L_x}\,. \label{eq37} \eeq
This is nothing but the deviation of the coordinate $y_0$ such that it
changes the quantum number $l$ by one. The generators of the quantum
algebra \eq{ED}-\eq{EF} behave on the representation basis
\eq{EJ} as
\beq
{\cal L}_{\pm1}\vert n,l\rangle
=[{1\over2}\pm l]\vert n,l\pm1\rangle\,,\label{EK}
\eeq
\beq
{\cal K}\vert n,l\rangle=q^{2l}\vert n,l\rangle\,.\label{EL} \eeq
${\cal K}$ measures the quantum number $l$ and ${\cal L}_{\pm1}$ raises
(lowers) $l$. We remark that our quantum algebra is associated with
only the quantum number $l$, namely the degeneracy of the Landau
levels. The energy level $n$ is invariant under the action of the
quantum algebra. This is the difference from the case of the
$su(2)$ algebra \eq{EX}.

Second, we discuss the case of a periodic potential in the same gauge
\beq
H={1\over2m}({\hat p}_1+m\omega y)^2+{1\over 2m}{\hat p}_2^2+V(x,y)
\label{eq40} \eeq
where
\[ V(x+a_1,y)=V(x,y+a_2)=V(x,y)\,. \]
According to the Bloch theorem, we put the wave function of the form
\beqa
& \psi(x,y)=exp({i\over\hbar}p_x x)\phi(x,y) \label{eq41} \\
& \phi(x+a_1,y)=\phi(x,y) \label{eq42} \eeqa
into the Schr{\"o}dinger equation $H\psi=E\psi$. It is convenient to
introduce the dimensionless quantities:
\beqa
&\xi=l_B^{-1}x\,,\hskip 20pt \eta=l_B^{-1}y+l_Bp_x\hbar^{-1}\,, \nn\\
&E={1\over2}\hbar\omega\epsilon\,,\hskip 20pt V(x,y)=m\omega\hbar
v(\xi,\eta) \label{eq43} \\
&u(\xi,\eta)=\phi(x,y+l_B^2p_x\hbar^{-1})\,. \nn
\eeqa
If $l_B^2p_x\hbar^{-1}$ is proportional to the period $a_2$ or $V$ is
independent of $y$, $H\psi=E\psi$ becomes
\beq
\left( ({d\over d\xi}+i\eta)^2+{d^2\over d\eta^2}+\epsilon
-v(\xi,\eta)\right) u(\xi,\eta)=0\,. \label{EP} \eeq
We see that $u$ will be solved as a function of only three
variables $\xi$, $\eta$ and $\epsilon$. We thus fix the form of the
eigenfunctions
\beq
\vert\epsilon,p\rangle\equiv exp({i\over\hbar}px)f(l_B^{-1}x,
l_B^{-1}y+l_Bp\hbar^{-1},\epsilon)\,. \label{EU} \eeq
We consider the case of the following degeneracy
\beqa
&\epsilon(p_1)=\epsilon(p_2)=\dots=\epsilon(p_{2j+1})\nn \\
&p_k=2\pi\hbar{n_k\over L_x}\hskip 20pt (k=1,\dots,2j+1)
\label{eq46} \eeqa
where $j$ is a positive integer and $n_k=-j+k-1$. The magnetic
translation acts on the state \eq{EU} as
\beq
T_{({\bar\Delta},\Delta)}\vert\epsilon,p_k\rangle=
exp\left({i\over\hbar}{\bar\Delta}({1\over2}m\omega\Delta+p_k)\right)
\vert\epsilon,p_{k+1}\rangle\,, \label{EO} \eeq
where ${\bar \Delta}$ is an unknown parameter which will be determined
later by the representation theory. We therefore find the quantum
algebra given by the following generators
\beq
{\cal L}_{1}={T_{({\bar\Delta},\Delta)}
          -T_{(-{\bar\Delta},\Delta)}\over q-q^{-1}} \label{EM} \eeq
\beq
{\cal L}_{-1}={T_{(-{\bar\Delta},-\Delta)}
          -T_{({\bar\Delta},-\Delta)}\over q-q^{-1}} \label{eq49} \eeq
\beq
{\cal K}=T_{(2{\bar\Delta},0)}. \label{eq50} \eeq
and
\beq
q=exp(il_B^{-2}\Delta{\bar\Delta})\,. \label{EN} \eeq
When
\beq
\Delta=N_2a_2\hskip 25pt{\bar\Delta}=N_1a_1 \,\,,\label{EV}\eeq
where $N_1$ and $N_2$ are integers, only the magnetic translation
operators still commute with the Hamiltonian and thus our quantum
algebra commutes with the Hamiltonian.
The action on the eigenstates are obtained using \eq{EO}
\beqa
&{\cal L}_{\pm1}\vert\epsilon,n_k\rangle
=[{1\over2}\pm n_k]\vert\epsilon,n_{k\pm1}\rangle\,, \nn \\
&{\cal K}\vert\epsilon,n_k\rangle=q^{2n_k}\vert\epsilon,n_k\rangle\,.
\label{ET} \eeqa

Now let us determine ${\bar\Delta}$. The number of the degenerate
states being $n=2j+1$, the dimension of the representation is $2j+1$.
We notice that our representation \eq{ET} coincides with the spin-$j$
representation of ${\cal U}_q(su(2))$ when $q^n=1$.
One of the allowed values of $q$ is thus
\beq
q=exp({2\pi i\over 2j+1}) \label{1993}
\eeq
and ${\bar\Delta}$ is determined
\beq
{\bar\Delta}={L_x\over 2j+1}\,.\label{eq54}
\eeq
The condition \eq{EV} is satisfied when $L_x$ and $L_y$ are written as
\beq
L_x=(2j+1)N_1a_1\,, \hskip 30pt L_y=(2j+1)N_2a_2\,. \label{eq81}
\eeq

Finally, some remarks are in order.
In eqs.\eq{EM}-\eq{EN}, we have introduced another parameter
${\bar\Delta}$ which discriminates the representations. Also in
eqs.\eq{ED}-\eq{EI}, it is of course possible to introduce
${\bar\Delta}$. However in this case, the number of the degenerate
states being given by $n=L_xL_y/2\pi l_B^2$, ${\bar\Delta}$ amounts to
$(L_x/L_y)\Delta$. When $L_x=L_y$, ${\bar\Delta}$ becomes $\Delta$
after all. While in the case of $V\not=0$ we notice that
${\bar\Delta}\not=\Delta$ even if $L_x=L_y$. \par

The limit $q\rightarrow1$ can be performed by the double
scaling limit $L_x\rightarrow\infty$ and $B\rightarrow0$ keeping
$L_xB=const.$ or the value $\Delta$ (for example, $L_xB=2\pi\hbar c/
e\Delta$). The generators ${\cal L}_{\pm1}$ and the RHS of the
commutation relation $({\cal K}-{\cal K}^{-1})/(q-q^{-1})$ become
singular on $l_B^{-2}\rightarrow0$. The singular part of the generators
is simply given by replacing $T_{\bf \alpha}$ with $\tau_{\bf \alpha}$.
Hence appropriately rescaled generators of the quantum algebra reduce
to the $u(1)$ generators, {\em i.e.}, just the translations when we
consider the limit $q\rightarrow1$. \par

The quantum group structure is found to be relevant to the degeneracies
of the Landau Level. So far as a periodic potential is concerned,
it is also known that each Landau level splits into some subbands of
equal weight under a weak sinusoidal potential \cite{Rauh}. We can
deduce that the representation basis of a quantum group is
composed of wavefunctions labeled by quantum numbers of these
degeneracies and that there exists a quantum group symmetry with the
value \eq{1993}.

We have assumed $l_B^2p_x\hbar^{-1} \propto a_2$ in deriving eq.
\eq{EP}. The proportional constant is the integer $N_2n_k$. This
condition implies that the magnetic-flux quanta (fluxon) per unit cell
$\phi_a =a_1a_2 eB/hc$ is given by
\beq
\phi_a={1 \over N_1N_2(2j+1)}\,. \label{91}
\eeq
The electron density per unit cell is
\beq
\rho_a={a_1a_2\over L_xL_y}={1 \over N_1N_2(2j+1)^2}\,, \label{92}
\eeq
and the filling factor is thus
\beq
\nu={\rho_a \over\phi_a}={1\over 2j+1}. \label{93}
\eeq
This is a fractional filling of $\nu=1/m$ ($m$ odd), which appears
in the Laughlin function \cite{Laugh}. It might be possible to examine
a many-particle system along the same line as this paper and we
therefore expect the existence of a quantum group symmetry with the
value of deformation parameter
\beq
q=exp(2\pi i \nu)\,.
\eeq
It is not until the quantum group symmetry exists that the above
relation will be proved. Furthermore, $q$ being a root of unity, the
representation theory \cite{Lu} is considerably different from those of
\eq{EX},\eq{eq14} and \eq{EA}. Hence, our derivation of the quantum
group symmetry will give some nontrivial suggestions.

Generalizing the consideration of this paper to many-particle
systems, we might obtain a new feature of fractional Hall systems and
of other condensed matter problems \cite{any} related to the
Landau-level degeneracies.

\vspace{1cm}
\noindent
{\em Acknowledgements}

The author would like to thank N. Aizawa, K. Higashijima, N. Ohta
and H. Shinke for valuable discussions and useful suggestions.

%
%
%

%
\end{document}